\documentclass[12pt]{article}
\usepackage{graphicx}
\usepackage{amsmath}
\usepackage{amssymb}

\begin{document}

\vskip1cm
\begin{center}
{\LARGE {\bf Holographic descriptions of (near-)extremal black holes in five dimensional minimal supergravity}}
\vskip3cm
{\large 
{
Wen-Yu Wen\footnote{e-mail address: steve.wen@gmail.com}
}
}

\vskip1cm

{\it Department of Physics and Center for Theoretical Sciences,\\
National Taiwan University, 
Taipei 106, TAIWAN}

\vskip0.5cm

{\it Leung Center for Cosmology and Particle Astrophysics,\\
National Taiwan University, 
Taipei 106, TAIWAN}

\end{center}

\vskip1cm
\begin{abstract}

We have learnt that the Bekenstein-Hawking entropy of supersymmetric black holes in five-dimensional minimal supergravity can be reproduced from microstate counting via the AdS/CFT and Kerr/CFT correspondence.  In this note, we demostrate the coincidence of both formalisms for extremal and near-extremal black holes in five-dimensional minimal supergravity.  Connection to $AdS_2$ gravity is also mentioned.

\end{abstract}


\vfill\eject

\section{Introduction and Summary}
It has been shown recently that the near-horizon geometry of four-dimensional Kerr black hole at its extremal limit possesses the isometry $SL(2,R)\times U(1)$, where a chiral CFT was formulated with appropriate boundary conditions and Bekenstein-Hawking entropy can be reproduced by the Cardy formula\cite{Guica:2008mu}, denoting as the Kerr/CFT correspondence.   A sample of earlier CFT appraoches to black hole entropy can be found in \cite{Strominger:1996sh,Maldacena:1997de,Carlip:1998wz,Solodukhin:1998tc,Park:2001zn,Loran:2008mm}.  The Kerr/CFT correspondence was later applied to other extremal black holes and relevant discussion can be found in \cite{Lu:2008jk,Azeyanagi:2008kb,Hartman:2008pb,Nakayama:2008kg,Chow:2008dp,Isono:2008kx,Azeyanagi:2008dk,Chen:2009xja,Peng:2009ty,Loran:2009cr,Ghezelbash:2009gf,Lu:2009gj,Compere:2009dp,Wu:2009cn,Hotta:2009bm,Astefanesei:2009sh,Garousi:2009zx,Ghezelbash:2009gy,Krishnan:2009tj}.  In particular, in \cite{Isono:2008kx} the authors investigated a class of black hole  which at the extremal limit is in fact supersymmetric and its near-horizon geometry is the near-horizon geometry of extremal BTZ black hole\cite{Banados:1992wn}.  A chiral CFT was found via the Kerr/CFT approach and the Bekenstein-Hawking entropy was correctly reproduced.  The isometry for this near-horizon geometry is in fact enhanced to $SO(2,2)$ and a non-chiral CFT has been identified at the asymptotic boundary of $AdS_3$ with appropiate boundary conditions\cite{Brown:1986nw}, which also gives rise to the correct entropy by the Cardy formula\cite{Strominger:1997eq}.  We will denote the latter approach as the usual AdS/CFT in this paper.  It is, however, unnecessary for those CFT's in both approaches of Kerr/CFT and AdS/CFT to be the same for their differences in chirality, while the boundary conditions imposed in order to obtain finite conserved charges are also quite different.  Nevertheless, both formalisms seem to be {\sl physically} identical in this class of supersymmetric black holes because they give the same central charge and oscillator level (or Frolov-Thorne temperature).  In this note, we explicitly show that the near-horizon geometry of supersymmetric black holes (BMPV) and black rings in five-dimensional minimal supergravity can be casted into a special form of $AdS_3$, where the isometry $SL(2,R)\times U(1)\in SO(2,2)$ is manifested.  Therefore either formalism of Kerr/CFT or AdS/CFT can be applied and the Bekenstein-Hawking entropy can be reproduced by the Cardy formula.  We further investigate the near-extremal limit where the supersymmetry is slightly broken, and have found again the coincidence of both formalisms.  In the extremal limit the non-chiral CFT given in the AdS/CFT approach is {\sl effectively} chiral due to the BPS condition, and one may suspect it is the very reason two CFT's agree.  In the near-extremal limit, however, we do have both left and right-moving sectors of CFT turned on and there is no ambiguity in its non-chirality.  Nevertheless, in the near-extremal limit we still have a well-defined chiral CFT in the Kerr/CFT approach, which reproduces desired entropy for near-extremal black holes.  By investigating the very near-horizon geoemtry at both outer and inner horizons, we are able to realize a non-chiral CFT dual to BTZ geometry from two copies of chiral CFT's dual to near-extremal Kerr metric at each horizon.  Upon dimensional reduction to $AdS_2$, we obtain a 2D gravity coupled to a dilaton and electric field, where entropy can also be obtained for a chiral CFT dual to this background.  This paper is organized as follows:  In section $2$, we review the construction of chiral CFT dual to the extremal Kerr metric of the supersymmetric black holes (rings) in five-dimensional minimal supergravity.  In section $3$, we review the construction of non-chiral CFT dual to the extremal BTZ metric, as a special choice of coefficients in a general extremal Kerr metric.  In section $4$, we discuss the common physical quantities but difference of boundary conditions between two approaches.  In section $5$, we derive the chiral CFT for near-extremal black holes, which possesses a non-extremal BTZ metric in the near-horizon geometry.  We also argue that the non-chiral CFT dual to the BTZ metric can be constructed from two copies of chiral CFT's dual to very near-outer and inner horizon geometry.  At last we comment on its relation to $AdS_2$ gravity.  For completeness, in the appendix A we highlight the derivation of near-horizon geometry of suprsymmetric black holes and rings in five-dimensional minimal supergravity.  In the appendix B we recall the coordinate transformation which brings near-horizon geometry to AdS$_3$ metric.  In the appendix C we review the derivation of non-extrmal BTZ metric as part the near-horizon geometry of some non-extremal black holes.  In the appendix D, we review the Kaluza-Klein reduction from $AdS_3$ to $AdS_2$.  Although in this note we only consider (near-)extremal black holes in five dimensions, but it can be easily generalized to higher dimensions as long as the BTZ metric is part of the near-horizon geometry.

\section{Kerr/CFT correspondence from AdS metric}
The supersymmetric solutions in the five-dimensional minimal supergravity have been classified by \cite{Gauntlett:2002nw}.  The black hole solution with horizon topology $S^3$ was first discovered by \cite{Breckenridge:1996is} and later the black ring solution with horizon topology $S^1\times S^2$ was constructed in \cite{Elvang:2004rt}.  Both solutions are imitatedly related and have string theory interpretation by the D$1$-D$5$-P system or M-theory interpretation by supertubes.  As reviewed in the appendix A, the near-horizon geometry of both supersymmetric black holes and black rings, regardless different horizon topologies, can be casted into the following metric:
\begin{equation}\label{AdS_metric}
ds^2=\frac{4r_+}{l}rdtd\psi+r_+^2d\psi^2+\frac{l^2}{4r^2}dr^2+\frac{l^2}{4}(d\theta^2+\sin^2{\theta}d\chi^2)
\end{equation}
which is a direct product of $AdS_3$ and a compact $S^2$.  Now we can rearrange it into another form where isometry $SL(2,R)\times U(1)$ is manifested by sending $t\to \frac{l^2}{4}t$, that is
\begin{equation}
ds^2=\frac{l^2}{4}(-r^2dt^2+\frac{dr^2}{r^2}+d\theta^2)+r_+^2(d\psi+\frac{l}{2r_+}rd\hat{t})^2+\frac{l^2}{4}\sin^2{\theta}d\chi^2
\end{equation}

To apply Kerr/CFT correspondence, we will choose the following boundary condition\cite{Guica:2008mu}:

\begin{align}
\left(
\begin{array}{ccccc}
h_{tt}=\mathcal{O}(r^2)
&h_{tr}=\mathcal{O}(\frac{1}{r^2})
&h_{t\theta}=\mathcal{O}(\frac{1}{r})
&h_{t\psi}=\mathcal{O}(1)
&h_{t\chi}=\mathcal{O}(r)\\
h_{rt}=h_{tr}
&h_{rr}=\mathcal{O}(\frac{1}{r^3})
&h_{r\theta}=\mathcal{O}(\frac{1}{r^2})
&h_{r\psi}=\mathcal{O}(\frac{1}{r})
&h_{r\chi}=\mathcal{O}(\frac{1}{r^2}) \\
h_{\theta t}=h_{t\theta}
&h_{\theta r}=h_{r\theta}
&h_{\theta\theta}=\mathcal{O}(\frac{1}{r})
&h_{\theta\psi}=\mathcal{O}(\frac{1}{r})
&h_{\theta \chi} = \mathcal{O}(\frac{1}{r}) \\
h_{\psi t}=h_{t\phi_1}
&h_{\psi r} =h_{r\phi_1}
&h_{\psi\theta} =h_{\theta\phi_1}
&h_{\psi\psi}=\mathcal{O}(1)
&h_{\psi \chi}=\mathcal{O}(1)\\
h_{\chi t}=h_{t\phi_2}
&h_{\chi r}= h_{r\phi_2}
&h_{\chi\theta}=h_{\theta \phi_2}
&h_{\chi\psi} =h_{\phi_1 \phi_2}
&h_{\chi\chi}=\mathcal{O}(\frac{1}{r})
\end{array}
\right),
\label{boundary_condition}
\end{align}

and the most general diffeomorphism which preserves 
this boundary condition reads
\begin{equation}
\zeta =
\Bigl[C+\mathcal{O}\bigl(\frac{1}{r^3}\bigr)\Bigr]\partial_t
+[-r\epsilon'(\psi)+\mathcal{O}(1)]\partial_{r}
+\mathcal{O}\bigl(\frac{1}{r}\bigr)\partial_{\theta} +\mathcal{O}\bigl(\frac{1}{r^2}\bigr)\partial_{\chi}
+\Bigl[\epsilon(\psi)+\mathcal{O}\bigl(\frac{1}{r^2}\bigr)\Bigr]\partial_\psi,
\end{equation}
where $C$ is an arbitrary constant and
$\epsilon(\psi)$ is an arbitrary function of $\psi$.
Therefore, the asymptotic symmetry group is generated by 
\begin{eqnarray}
&&\zeta^t=\partial_{t}, \label{time_killing}\\
&&\zeta^{\psi}=\epsilon(\psi)\partial_{\psi}-r\epsilon'(\psi)\partial_r.
\end{eqnarray}

To compute the central charge on dual CFT side, we choose the Fourier components $\epsilon_n(\psi)=e^{-in\psi}$ and 
the commutator of $\zeta_{(n)}^{\psi}$ constitutes 
a copy of chiral Virasoro algebra with zero central charge.
The central extension $c$ of Virasoro algebra is given as follows\cite{Barnich:2001jy,Barnich:2007bf}:
\begin{eqnarray}\label{c}
&{}&
\frac{1}{8\pi G_5}\int_{\partial\Sigma}k_{\zeta_{(m)}^{[1]}}[\mathcal{L}_{\zeta_{(n)}^{[1]}}g, g]=
-\frac{i}{12}(m^3+xm)\delta_{m+n}c,
\end{eqnarray}
where $\partial\Sigma$ is a spatial slice. The 3-form $k_{\zeta}$ is defined by
\begin{eqnarray}
&{}&
k_{\zeta}[h,g]=
\frac{1}{2}\left[\zeta_{\nu}D_{\mu}h-\zeta_{\nu}D_{\sigma}h_{\mu}^{~\sigma}+\zeta_{\sigma}D_{\nu}h_{\mu}^{~\sigma}
+\frac{1}{2}hD_{\nu}\zeta_{\mu}-h_{\nu}^{~\sigma}D_{\sigma}\zeta_{\mu}
+\frac{1}{2}h^{\sigma}_{~\nu}(D_{\mu}\zeta_{\sigma}+D_{\sigma}\zeta_{\mu})\right] \nonumber\\
&{}&~~~~~~~~~~~~*(dx^{\mu}\wedge dx^{\nu}),
\label{cc}
\end{eqnarray}  
where $g$ denotes the near-horizon metric.
The coefficient $x$ in \eqref{c} is unimportant because it can be absorbed by a shift of $L_0$.  Here we obtain 
\begin{equation}
c=\frac{3\pi l^3}{2G_5}, \qquad T_{FT}=\frac{1}{2\pi (l/2r_+)}=\frac{r_+}{\pi l}, 
\end{equation}
for central charge and Frolov-Thorne temperature.  Therefore the Bekenstein-Hawking entropy can be reproduced by the Cardy formula, 
\begin{equation}\label{eq:cardy1}
S=\frac{\pi^2}{3}cT_{FT} = \frac{(2\pi r_+)(\pi l^2)}{4G_5}
\end{equation}

\section{AdS/CFT correspondence from Kerr metric}
The near-horizon geomety used in Kerr/CFT correspondence, denotating as Kerr metric, takes the following general form\cite{Kunduri:2007vf}:
\begin{equation}\label{Kerr_metric}
ds^2 = \Gamma(\theta) [-r^2 dt^2+\frac{dr^2}{r^2}+\alpha(\theta)d\theta^2]+\gamma_1(\theta)(d\phi_1+krdt)^2+\gamma_2(\theta)d\phi_2^2
\end{equation}
where we intend to keep only one spin along $\phi_1$ for simplicity.  It has been proved in \cite{Reall:2002bh} that a supersymmetric solution with a compact horizon has a horizon geometry of $T^3$, $S^1\times S^2$ or (squashed) $S^3$.  We are interested in the latter two geometries, which correspond to the black ring and BMPV black hole respectively.  To have desired near-horizon geometry, by direct computation, we learn that all the undetermined functions in metric components are in fact constant functions except $\gamma_2(\theta)=\Gamma\alpha \sin^2{\theta}$.  Moreover, the condition for existence of a Killing spinor necessiates a near-horizon geometry $AdS_3$ spanned by coordinates $(t,r,\phi_1)$, which in turn requires $\Gamma=k^2\gamma_1$.  As a result, the metric becomes
\begin{equation}
ds^2=2\sqrt{\gamma_1\Gamma}rdtd\phi_1+\gamma_1d\phi_1^2+\frac{\Gamma}{r^2}dr^2+\Gamma(d\theta^2+\sin^2{\theta}d\phi_2^2)
\end{equation}
The metric is a product space of $AdS_3$ of curvature radius $\ell=2\sqrt{\Gamma}$ and a $S^2$ of radius $\sqrt{\Gamma}$.  Identifying the $AdS_3$ as near-horizon geometry of extremal BTZ black hole, using AdS/CFT correspondence with boundary condition given in \cite{Brown:1986nw}, one obtains the central charge
\begin{equation}
\hat{c}=\frac{3\ell}{2G_3}=\frac{12\pi}{G_5}\Gamma^{3/2}
\end{equation}
where $G_3=G_5/(4\pi\Gamma)$.  To use the Cardy formula, we also have to know the oscillator level $q_0$, which is the eigenvalue to the Virasoro gnerator $L_0$.  A method to obtain it is from the Komar integral at the horizon\cite{Emparan:2005jk}.  We first define a one-form $v$ dual to the Killing vector field $\partial_{\phi_1}$ by
\begin{equation}
v = \gamma_1 d\phi_1 + \sqrt{\gamma_1\Gamma}rdt.
\end{equation}
Then the Komar integral evaluated at the horizon is 
\begin{equation}
q_0=\frac{1}{16\pi G_5}\int_{S^1\times S^2}{*dv}=\frac{\pi}{2G_5}\gamma_1\sqrt{\Gamma}.
\end{equation}
Therefore, the Cardy formula
\begin{equation}\label{eq:cardy2}
S = 2\pi \sqrt{\frac{\hat{c}q_0}{6}} = \frac{(2\pi\sqrt{\gamma_1})(4\pi\Gamma)}{4G_5}
\end{equation}
reproduces the Bekenstein-Hawking entropy.

\section{Boundary conditions}
We have demonstated that both AdS/CFT correspondence and Kerr/CFT correspondence can provide a microscopic derivation for the entropy of five-dimensional supersymmetric black hole (ring).  Similar argument can be applied to supersymmetric balck holes in arbitrary higher dimensions as long as a typical form of $AdS_3$ metric can be factored out in the near-horizon geometry.   Recognizing it as the near-horizon of extremal BTZ black hole, one is able to identify a non-chiral CFT by imposing boundary conditions a la Brown-Henneaux \cite{Brown:1986nw} on one hand but a chiral CFT by imposing boundary conditions a la Guica-Hartman-Son-Strominger \cite{Guica:2008mu} on the other hand.  It is tempting to make the following identification:
\begin{equation}
T_{FT} = (\frac{\partial S}{\partial q_0})^{-1},\qquad c=\hat{c}.
\end{equation}
where the oscillator level is given by the angular momentum at the horizon, implied by the Komar integral, and central charges are identified in each CFT.  Then the Cardy formula given by (\ref{eq:cardy1}) and (\ref{eq:cardy2}) are simply related by the Legendre transformation of the thermodynamics first law.  Although effectively only the left-moving sector contributes to entropy in the non-chiral CFT dual to AdS metric thanks to its extremality, one has to be cautious that the boundary conditions could be different from that in the chiral CFT dual to Kerr metric.  In fact, the coordinate transformation which brings AdS$_3$ to near-horizon of extremal BTZ implies that the boundary conditions imposed by Brown-Henneaux is weaker, that is,
\begin{align}
\left(
\begin{array}{cccc}
h_{tt}=\mathcal{O}(1)
&h_{tr}=\mathcal{O}(\frac{1}{r^2})
&h_{t\phi_1}=\mathcal{O}(1)\\
h_{rt}=h_{tr}
&h_{rr}=\mathcal{O}(\frac{1}{r^3})
&h_{r\phi_1}=\mathcal{O}(\frac{1}{r^2})
&\cdots\\
h_{\phi_1 t}=h_{t\phi_1}
&h_{\phi_1 r} =h_{r\phi_1}
&h_{\phi_1\phi_1}=\mathcal{O}(1)\\
&\vdots&&\ddots
\end{array}
\right).
\label{boundary_condition}
\end{align}
As long as the most general diffeomorphism is concerned, this may not be able to produce a nontrivial central charge for the chiral CFT dual to Kerr metric.

\section{Holographic descriptions of near-extremal black holes}
\subsection{CFT dual to Kerr metric}
We have learnt that the AdS/CFT correspondence can also be applied to some near-extremal black holes whose extremal limit become supersymmetric.  In this case, both left and right CFT's contribute to the entropy.  Therefore it is well motivated and nontrivial to ask whether the chiral CFT description can still be valid  away from the extremality.  We will start with the near-horizon geometry\footnote{We remark that here near-horizon geometry refers to the near $M5$ branes goemetry as detailed in the Appendix C, to be distinguished from the near-horizon geometry of black hole.} of some near-extremal black hole which contains a non-extremal BTZ and $S^2$:
\begin{equation}
ds^2=-\frac{(r^2-r_+^2)(r^2-r_-^2)}{l^2r^2}dt^2+\frac{l^2r^2}{(r^2-r_+^2)(r^2-r_-^2)}dr^2+r^2(d\phi-\frac{r_+r_-}{lr^2}dt)^2+\frac{l^2}{4}(d\theta^2+\sin^2{\theta}d\psi^2)
\end{equation} 
Upon zooming in the region just ouside the outer horizon, say $r\to r_++\epsilon u, t \to \frac{l^2t}{4\epsilon}$ and at the same time keeping the near-extremal limit $T_{H}\to \epsilon T_{H}$, one obtains the near-outer-horizon geometry by sending $\epsilon\to 0$,
\begin{eqnarray}\label{NEBTZ}
&&ds^2 = \frac{l^2}{4}(-u(u+\Delta)dt^2 + \frac{du^2}{u(u+\Delta)}) + r_+^2(d\tilde{\phi}+\frac{lr_-}{2r_+^2}udt)^2+\frac{l^2}{4}(d\theta^2+\sin^2{\theta}d\psi^2),\nonumber\\
&&\Delta \equiv \frac{\pi l^3 T_H}{2r_+},
\end{eqnarray}
where we have defined $\tilde{\phi}\equiv \phi - \frac{r_-}{r_+}\frac{t}{l}$.  We remark that the first two terms can be brought into a standard $AdS_2$ metric\cite{Spradlin:1999bn}, however it is sufficient for us that the isometry $SL(2,R)\times U(1)$ manifests in the asymptotic geometry as $u\to \infty$ thanks to the subleading $\Delta\sim {\cal O}(1)$.  Therefore we expect the same asymptotic symmetry group and diffeomorphism as those for extremal Kerr black holes.  We then apply Kerr/CFT correspondence to compute the central charge and Frolov-Thorne temperature as follows:
\begin{equation}
c_+=\frac{3\pi l^3 r_-}{2G_5r_+}, \qquad T_+=\frac{r_+^2}{\pi l r_-}.
\end{equation}
and the Bekenstein-Hawking entropy is reporduced,
\begin{equation}
S_+=\frac{\pi^2}{3}c_+T_+ = \frac{(2\pi r_+)(\pi l^2)}{4G_5}.
\end{equation}
\subsection{CFT dual to BTZ metric}
To make connection to the non-chiral CFT dual to BTZ metric, one can also zoom in the region just inside the inner horizon\footnote{We remark that this region shares the same spacetime signature as the region outside the outer horizon and we expect the quantum vacuum is also well defined.} and obtain another chiral CFT with central charge and Frolov-Thorne temperature
\begin{equation}
c_-=\frac{3\pi l^3 r_+}{2G_5 r_-}, \qquad T_-=\frac{r_-^2}{\pi l r_+}.
\end{equation}
In this way, we can obtain a second entropy associated with inner horizon area,
\begin{equation}
S_-=\frac{\pi^2}{3}c_-T_- = \frac{(2\pi r_-)(\pi l^2)}{4G_5}.
\end{equation}
Together, the entropies of both left and right sectors of non-chiral CFT are related as follows:
\begin{equation}
S_L=\frac{1}{2}(S_+ + S_-), \qquad S_R=\frac{1}{2}(S_+ - S_-).
\end{equation}
It appears that the entropy $S_{R,L}$ are generated through entanglement between two qunatum vacua at outer and inner horizons.  A relevant discussion about entanglement entropy between conformal qunatum mechanics living at two boundaries of $AdS_2$ has been given in \cite{Azeyanagi:2007bj}.

\subsection{CFT dual to $AdS_2$ gravity}
A chiral CFT can also be constructed dual to $AdS_2$ gravity, and the ground state entropy of CFT is responsible to the entropy of $AdS_2$ black hole.  Here we demonstrate its relation to 5D near-extremal black hole.  If we apply dimensional reduction as detailed in the appendix D to the metric (\ref{NEBTZ}), forgetting the $S^2$ part for the moment, we obtain a $AdS_2$ metric, dilaton field $\Phi$, and electric field $\tilde{F}_{tu}$:
\begin{eqnarray}
&&\tilde{ds}^2 = \frac{l^2}{4}[-u(u+\Delta)dt^2+\frac{du^2}{u(u+\Delta)}],\nonumber\\
&&e^{2\Phi} = \frac{l^2}{r_+^2},\nonumber\\
&&\tilde{F}_{tu} = \frac{lr_-}{2r_+^2}.
\end{eqnarray}
The 2D central charge which agrees with the Brown-Henneaux 3D central charge reads\footnote{We have chosen a different normalization for Newton constant, say $2\pi r_+ G_2 = G_3$, from that in \cite{Castro:2008ms}}, 
\begin{equation}
c_2 = \frac{3}{2G_2}.
\end{equation}
It was suggested in \cite{Castro:2008ms} that a 2D chiral CFT with degrees of freedom $c_0$ and oscillator level $q_0$ given by 
\begin{equation}
c_0 = \frac{c_2}{\pi}, \qquad q_0 = \frac{c_0}{24}
\end{equation} 
will reproduce Bekenstein-Hawking entropy
\begin{equation}
S = 2\pi \sqrt{\frac{c_0 q_0}{6}} = \frac{1}{4G_2}.
\end{equation}
Now, recollecting the normalization of Newton constant in various dimensions and putting together with $S^2$, we have
\begin{equation}
S = \frac{1}{4G_2} = \frac{2\pi r_+}{4G_3} = \frac{(2\pi r_+)(\pi l^2)}{4 G_5}
\end{equation}

\subsection{Comments}
We have learnt that near-horizon metric (\ref{NEBTZ}) is responsible for 5D near-extremal black hole where supersymmetry is slightly broken.  It is not difficult to see that a general form similar to (\ref{Kerr_metric}) can be obtained from near-horizon geometry of near-extremal Kerr(-Newman) black holes in arbitrary dimensions $D\ge 4$, that is
\begin{equation}
ds^2 = \Gamma(\theta) [-r(r+\Delta) dt^2+\frac{dr^2}{r(r+\Delta)}+\alpha(\theta)d\theta^2]+\sum_{i=1}^{D-3}{\gamma_i(\theta)(d\phi_i+k_irdt)^2},
\end{equation}
for some finite $\Delta$.  Then one may claim there exist a set of chiral CFT's dual to this metric and each possesses central charge and Frolov-Thorne temperature as follows:
\begin{equation}
c_i=3k_i \int_0^{\pi}{d\theta\sqrt{\Gamma(\theta)\alpha(\theta)\gamma_i(\theta)}},\qquad T_i = \frac{1}{2\pi k_i},
\end{equation}
such that the Bekenstein-Hawking entropy can be reproduced by the Cardy formula as
\begin{equation}
S = \frac{\pi^2}{3}c_iT_i=\frac{\pi}{2}\int_0^{\pi}{d\theta\sqrt{\Gamma(\theta)\alpha(\theta)\gamma_i(\theta)}}.
\end{equation}
In this way, the Kerr/CFT correspondence can be carried over to the near-extremal limit without much modification.

\section*{Acknowledgements}
The author is grateful to Chiang-Mei Chen, Pei-Ming Ho, Hirotaka Irie, Shoichi Kawamoto, Furuuchi Kazuyuki, Darren Shih, Tomohisa Takimi, and Ta-Sheng Tai for useful comments; special thanks to Hiroshi Isono, for useful discussion and contribution in the early stage of this project.  This work is supported by the Taiwan's National Science Council and National Center for Theoretical Sciences under Grant No. NSC 97-2119-M-002-001 and 97-2112-M-002-015-MY3.

\begin{appendix}
\section{Near-horizon geometry of supersymmetric black holes in five-dimensional minimal supergravity}
In this section, we show that near-horizon of supersymmetric black holes can be brought into a desired form (\ref{AdS_metric}).
\subsection{BMPV black holes}
We start with the black hole metric\cite{Breckenridge:1996is}
\begin{eqnarray}
&&ds^2=-(1+\frac{l^2}{\rho^2})^{-2}[dt'+\frac{J}{2\rho^2}(d\psi'+\cos{\theta}d\chi)]^2+(1+\frac{l^2}{\rho^2})(d\rho^2+\rho^2d\Omega_3^2),\nonumber\\
&&d\Omega_3^2=\frac{1}{4}[d\theta^2+\sin^2{\theta}d\chi^2+(d\psi'+\cos{\theta}d\chi)^2].
\end{eqnarray}
The near-horizon metric is obtained as follows by taking $\rho\to 0$,
\begin{equation}
ds^2 = -[\frac{\rho^2}{l^2}dt'+\frac{J}{2l^2}(d\psi'+\cos{\theta}d\chi)]^2+l^2\frac{d\rho^2}{\rho^2}+l^2d\Omega_3^2
\end{equation}
To bring it into the desired form, we make the following coordinate transformation,
\begin{eqnarray}
&&r=\rho^2,\nonumber\\
&&d\psi = d\psi' - \Omega dt,\qquad \Omega \equiv \frac{r^2/l+Jr^2/l^4}{l^2/2-J^2/(2l^4)},\nonumber\\
&&t = \frac{l^2}{2\sqrt{l^6-J^2}}t'.
\end{eqnarray}
then we can obtain metric (\ref{AdS_metric}) by identifying $r_+^2 = \frac{l^2}{4}(1-\frac{J^2}{l^6})$.

\subsection{Black Rings}
We start with the black ring metric\cite{Elvang:2004rt}:
\begin{eqnarray}
&&ds^2 = -f^2 (dt+\omega)+ f^{-1}[d\rho^2+\rho^2\sin^2{\theta}d\phi^2+\rho^2\cos^2{\theta}d\psi^2],\nonumber\\
&& f^{-1}= 1+\frac{Q-q^2}{\Sigma}+\frac{q^2\rho^2}{\Sigma^2},\qquad \Sigma \equiv \sqrt{(\rho^2-R^2)^2+4R^2\rho^2\cos^2{\theta}},
\end{eqnarray}
where 
\begin{eqnarray}
&&\omega = \omega_\phi d\phi + \omega_\psi d\psi,\nonumber\\
&&\omega_\phi = -\frac{q\rho^2\cos^2{\theta}}{2\Sigma^2}[3Q-q^2(3-\frac{2\rho^2}{\Sigma})],\nonumber\\ 
&&\omega_\psi = -\frac{6qR^2\rho^2\sin^2{\theta}}{\Sigma(\rho^2+R^2+\Sigma)}- \frac{q\rho^2\sin^2{\theta}}{2\Sigma^2}[3Q-q^2(3-\frac{2\rho^2}{\Sigma})].
\end{eqnarray}
Following the discussion in \cite{Elvang:2004rt,Elvang:2004ds}, the near-horizon geoemtry can be identified as the near-horizon geoemtry of extremal BTZ black hole with mass $m_{BTZ}=2L^2/q^2$ and spin $J_{BTZ}=qM_{BTZ}$ as follows:
\begin{eqnarray}
&&ds^2 = \frac{q^2}{4}\frac{dr^2}{r^2} + \frac{4L}{q}rdt d\psi' + L^2 d\psi'^2+\frac{q^2}{4}(d\theta^2+\sin^2{\theta}d\chi^2),\nonumber\\
&&L\equiv \sqrt{3[\frac{(Q-q^2)^2}{4q^2}-R^2]},
\end{eqnarray}
which is the very metric (\ref{AdS_metric}) with $q=l$ and $r_+=L$.

\section{Near-horizon geometry of near-horizon extremal BTZ and AdS$_3$ goemetry}
In this section, we show the coordinate transformation between near-hirozon extremal BTZ and AdS$_3$ in Poincar\`{e} patch.  In this note, we have been paying attention to the near-horizon geometry which features a $AdS_3$ metric as follows:
\begin{equation}
ds^2=\frac{4r_+}{l}rdtd\psi + r_+^2 d\psi^2 +\frac{l^2}{4r^2}dr^2.
\end{equation}
This can be brought into a more familiar form of $AdS_3$ in the Poincar\`{e} patch locally:
\begin{equation}
ds^2=-\frac{\rho^2}{l^2}d\tau^2 + \frac{l^2}{\rho^2}d\rho^2 + \frac{\rho^2}{l^2}d\phi^2,
\end{equation}
by the following coordinate transformation
\begin{eqnarray}
&&\phi+\tau = \frac{l}{2r_+}e^{\frac{2r_+}{l}(\frac{t}{l}+\psi)},\nonumber\\
&&\phi-\tau = \psi - \frac{t}{l} -\frac{l}{2x},\nonumber\\
&&\frac{l}{\rho} = \frac{1}{\sqrt{2r_+ x}}e^{\frac{r_+}{l}(\frac{t}{l}+\psi)}
\end{eqnarray}

\section{Near-extremal black holes}
Here we review a typical class of near-extremal charged black holes in four dimensions, which can be lifted up to five dimensions as a black string and entropy can be computed from the non-chiral CFT dual to the BTZ metric as part of its near-horizon geometry\cite{Balasubramanian:1998ee}.  The four dimensional black hole, originally constructed from three intersecting M5-branes over a common direction that carries momnetum, has the following metric:
\begin{eqnarray}
&&ds^2=-(H_0H_1H_2H_3)^{-\frac{1}{2}}(1-\frac{r_0}{r})dt^2 + (H_0H_1H_2H_3)^{\frac{1}{2}}[\frac{1}{1-\frac{r_0}{r}}dr^2+r^2d\Omega_2^2],\nonumber\\
&&H_i \equiv 1+\frac{r_0\sinh^2{\delta_i}}{r}.
\end{eqnarray}
Upon oxidating the above metric with a compact circular direction $\phi$ of radius $R$, together with a coordinate transformation $\tau=tl/R$ and $\rho^2=2R^2(r+r_0\sinh^2{\delta_0})/l$, where $l\equiv r_0(\sinh{2\delta_1}\sinh{2\delta_2}\sinh{2\delta_3})^{1/3}$.  In the near brane region, say $r \ll r_0\sinh^2{\delta_i}$, we obtain 
\begin{eqnarray}
&&ds^2=-N^2 d\tau^2 + N^{-2} d\rho^2 + \rho^2 (d\phi + N_\phi d\tau)^2 +\frac{l^2}{4}d\Omega_2^2,\nonumber\\
&&N^2 \equiv \frac{\rho^2}{l^2}-\frac{2R^2r_0\cosh{2\delta_0}}{l^3}+\frac{R^4r_0^2\sinh^2{2\delta_0}}{l^4\rho^2}, \qquad N_{\phi}\equiv \frac{r_0R^2\sinh{2\delta_0}}{l^2\rho^2}.
\end{eqnarray}
We can easily identify it as non-extremal BTZ$\times S^2$.  A similar discussion of a family of four-dimensional near-extremal rotating black holes is also available in \cite{Cvetic:1999ja}.  The BTZ goemetry can also be found in higher-dimensional black holes, see \cite{Satoh:1998sg} for example.

\section{Dimensional reduction from $AdS_3$ to $AdS_2$ gravity}
We start with a pure three-dimensional gravity with a cosmological constant:
\begin{equation}
I_3 = \frac{1}{16\pi G_3}\int{d^3x}\sqrt{g}(R+\frac{2}{l^2})+\cdots,
\end{equation}
where we have omitted the boundary terms which are irrelevant to our discussion, but see \cite{Castro:2008ms} for detail.  Given a 3D metric for Kaluza-Klein reduction,
\begin{equation}
ds^2 = \tilde{g}_{\mu\nu}dx^\mu dx^\nu + e^{-2\Phi}l^2(d\psi+\tilde{A}_\mu dx^{\mu})^2,
\end{equation}
where the 2D metric $\tilde{g}_{\mu\nu}$, dilaton field $\Phi$ and gauge field $\tilde{A}_{\mu}$ are independent of compactified direction $\psi$.  We obtain the two-dimensional effective action:
\begin{equation}
\tilde{I}_2 = \frac{l}{8G_3}\int{d^2x}\sqrt{-\tilde{g}}e^{-\Phi} (\tilde{R}+\frac{2}{l^2}-\frac{l^2}{4}e^{-2\Phi}\tilde{F}^2).
\end{equation}

\end{appendix}

\end{document}